\documentclass[preprint,12pt]{aastex}

\begin{document}

\title{A Note on ``Probing the existence of the $E_{peak}$-$E_{iso}$
correlation in long Gamma Ray Bursts'' by Ghirlanda et al.  -
astro-ph/0502186.}
\author{Ehud Nakar\altaffilmark{1} \& Tsvi Piran\altaffilmark{1,}\altaffilmark{2}}

\altaffiltext{1} {Theoretical Astrophysics, Caltech, Pasadena, CA
91125 USA} \altaffiltext{2}{Racah Institute for Physics, The
Hebrew University, Jerusalem 91904, Israel}

\begin{abstract}
In Nakar \& Piran (2004) we have  shown  that a large fraction of
BATSE bursts cannot satisfy the Amati relation between the isotropic
energy - peak energy of GRBs. Our results have been corroborated by
a further analysis carried out by Band and Preece (2005). Recently
Ghirlanda et al. (2005) claimed that when the exact distribution of
the scatter around the Amati relation is taken into account, the
BATSE sample is consistent with the Amati relation.  We show that
the probability that the BATSE sample is consistent with the Amati
relation (including the scattering), as found by Ghirlanda et al.
(2005) for a sample of 23 bursts with known redshift, is vanishingly
small.
\end{abstract}

Amati et al., (2002) suggested a tight relation between the
isotropic equivalent energy and the peak energy within GRBs (see
also Lamb et al., 2004; Atteia et al., 2004). This possible relation
between the intrinsic quantities of GRBs has generated a lot of
interest in view of its implication on the way GRBs operate and on
its potential applications. However this relation have been tested
so far only for a small sub-sample of GRBs with a known redshift.
Recently we (Nakar \& Piran,  2004) have questioned this
relationship and have shown that it cannot be satisfied by a large
fraction of the bursts in the much larger BATSE sample. Our results
have been supported by Band and Preece (2005) who performed a
similar test on a larger BATSE sample.

In a recent paper Ghirlanda et al. 2005 (GGF05) repeated a similar
test to the one preformed in Nakar \& Piran (2004) and Band \&
Preece (2005) with respect  the Amati relation found from a sample
of 23 bursts (their Eq. 1). This test checks the possibility that a
sample of bursts for which the redshifts are unknown is {\it
potentially consistent}\footnote{Note that while a burst that does
not satisfy the test cannot satisfy the Amati relation a burst that
satisfied this test does not necessarily satisfy the relation (see
Nakar \& Piran, 2004).} with the Amati relation. GGF05 claim that
they find that a large sample of 546 BATSE bursts is consistent with
the Amati relation. GGF05 argue that the difference between their
result and the results of Nakar \& Piran, (2004) and Band and Preece
(2005) arises from (i) their more accurate consideration of the
distribution of the dispersion around the Amati relation and (ii)
from  their requirement that a burst has to deviate by more than $3
\sigma$ from the relation in order to be considered as an outlier to
the relation.

We use here the BATSE sample as analyzed by GGF05 at face value, and
test the null hypothesis that this sample was drawn from the
distribution of the Amati relation as found for bursts with {\it
known} redshift by GGF05 (their Eq. 1). This relation is similar
(within $1\sigma$) to the original relation presented by Amati et
al. (2002) and later by Atteia et al., (2004). It is also similar to
the relation that is commonly used and that was tested by Nakar \&
Piran, (2004) and Band and Preece (2005). We find that the
probability that this null hypothesis is correct is vanishingly
small.

GGF05 express the Amati relation for bursts with measured
redshift(their Eq. 1) as:
\begin{equation}
\frac{E_p}{100 ~{\rm keV}}= (2.6 \pm 0.1)\left(\frac{E_{iso}}{1.17
\times 10^{53} ~{\rm erg}}\right)^\eta, \label{amati}
\end{equation}
where $\eta={0.56 \pm 0.02}$, $E_{iso}$  is the isotropic equivalent
energy and $E_p$ is the peak energy of the photon's spectral
distribution averaged over the whole burst. This relation was
derived by using the sample of 23 bursts with known redshifts
presented in Ghirlanda, Ghisellini and Lazzati (2004). They find
that the scatter of these bursts around Eq. 1 (in the log-log plane)
is best fitted by gaussian distribution with $\mu=0.1$ and
$\sigma=0.2$.  Hereafter we use the term Amati relation to describe
this distribution.

In order to find the minimal deviation of a burst with an unknown
redshift from the Amati relation, GGF05  reformulate Eq. 1 using the
observed bolometric flux, $F$ and observed peak energy $E_{p,obs}$
(Eq. 9 of GGF05):
\begin{equation}
\frac{(E_{p,obs})^{1/\eta}}{F}  = K^{1/\eta}  10^{\Delta/\eta} C(z),
\end{equation}
\begin{equation}
C(z) = \frac{4\pi d_L^2(z)}{(1 + z)^{(1+\eta)/\eta}}~.
\end{equation}
$d_L$ is the luminosity distance, $K=E_p/E_{iso}^{\eta}$ is given by
the rest frame Amati relation. The value of $\Delta$ for which Eq. 2
is satisfied defines the deviation of the bursts from Eq .1. Finding
the value of $\Delta$, requires a redshift. However, a lower limit
on $\Delta$ can still be found when the redshift is unknown,  using
the fact that $C(z)$ gets a maximum at $z = 5.3$. By replacing
$C(z)$ by $C_{max}=C(z=5.3)$ in Eq. 2, GGF05 obtain a lower limit on
$\Delta$ for a sample of 546 BATSE bursts. This procedure is similar
to the one carried by Nakar \& Piran (2004) and by Band \& Preece
(2005).

GGF05 presents the minimal fraction of bursts with a value of
$\Delta/\sigma$ that is larger than 0,1,2,3 in the first row of
table 1 of their paper. We show these values in the second column
of Table 1 below:

\begin{tabular}{|c|c|c|c|}
  \hline
  $\Delta/\sigma$ & Observed fraction  & Expected fraction  &Probability
  to find  \\
   & with a larger $\Delta/\sigma$  & with a larger $\Delta/\sigma$ & the observed fraction \\
  \hline
  0 & $84\%$ & $69\%$ & $10^{-15}$\\  \hline
  1 & $58\%$& $30\%$ &$10^{-38}$\\  \hline
  2 & $27\%$ & $6.7\%$ &$10^{-48}$\\  \hline
  3 & $7.7\%$ & $0.62\%$ &$10^{-31}$\\
  \hline
\end{tabular}

We test the null hypothesis that the BATSE sample was drawn from the
distribution of the Amati relation that GGF05 find for the 23 bursts
with known redshifts. According to this null hypothesis $\Delta$
should be distributed normally with $\mu=0.1$ and $\sigma=0.2$. In
table one we show the predictions of this hypothesis and the
probabilities that it is correct given the data. For example if the
BATSE sample was drawn from the Amati relation, as defined by GGF05,
only $69\%$ of the bursts should have $\Delta/\sigma>0$ in reality
at least $84\%$ of the observed BATSE bursts have $\Delta/\sigma>0$.
The probability to find this fraction under the null hypothesis is
$10^{-15}$. Similar tests can be carried out for bursts that are
have $\Delta/\sigma$ larger than 1,2 and 3 all give similar results.
Thus, using results of GGF05 (as they are) we find again (as we
found in Nakar \& Piran 2004)  that the BATSE sample is {\it
inconsistent} with the Amati relation as arise from the bursts with
known redshift.

We stress that the redshifts of these bursts are unknown. The
above measure for the minimal deviation from the Amati relation
makes the conservative assumption that the bursts are at the
optimal redshift for satisfying the Amati relation (for Eq. 2 this
redshift is $z = 5.3$). For bursts with a different redshift the
deviation would be larger. In other words the fraction in the
second column of our Table 1. is only a lower limit! The actual
discrepancy is even larger. A realistic redshift distribution
(Band \& Preece 2005) would make the discrepancy between the BATSE
sample and Eq. 1 even more prominent.

Ghirlanda et al. 2005 (GGF05) also consider  another version of
the Amati relation (GGF05 Eq. 6) that they derive using 442 BASTE
bursts with pseudo redshifts obtained using the lag-luminosity
relation. However,  Eq. 1. of GGF05 and  Eq. 6 of GGF05 are
inconsistent with each other. For example at $E_{iso} = 10^{52}$
erg they differ by $2.5\sigma$. Therefore the bursts with known
redshifts are inconsistent with the Amati relation as presented in
Eq. 6. This result demonstrates again that bursts with known
redshifts are not a representative sample of the BATSE bursts in
the sense that they cannot be used to draw a quantitative relation
between $E_{iso}$ and $E_p$.

GGF05 stress that the BATSE bursts are consistent (using our test)
with their Eq. 6. However,  as discussed earlier, our test  checks
for a given burst whether there is a redshift value for which Eq.
6 is satisfied. Since GGF05 have assigned pseudo redshifts to
BATSE bursts in order to derive Eq. 6 it is inevitable that BATSE
bursts will pass the test with this equation. Even thought the
number of tested bursts (546) is slightly larger than the number
of bursts used to derive Eq. 6 (442) this is not a consistency
check, this is a tautology.

Based on the data analysis of GGF05 we conclude (similarly to our
conclusion in Nakar \& Piran, (2004) and in agreement with the
findings of Band and Preece (2005)) that BATSE bursts do not
satisfy the  Amati relation  (Eq. \ref{amati}). Putting it
differently, while $E_{iso}$ and $E_p$ might be correlated, the
quantitative representation of such a correlation as given by the
Amati relation (based on the sample of bursts with a known
redshift) is inconsistent with the larger BATSE sample.

We thank G. Ghirlanda, G. Ghisellini, C. Firmani, D. Band and R.
Preece for helpful discussions.


\begin{thebibliography}{1}
\bibitem{Amatietal} Amati, L., Frontera, F., Tavani, M., in't Zand, J. J. M.,
Antonelli, A. et al.  Astr. Astrophys. 390, 81–89 (2002).
\bibitem[Atteia et al., 2004]{Atteiaetal} Atteia, J.-L., Ricker,
G.~R., Lamb, D.~Q., Sakamoto, T., Graziani, C., Donaghy, T.,
Barraud, C., \& The Hete-2 Science Team 2004, AIP Conf.~Proc.~727:
Gamma-Ray Bursts: 30 Years of Discovery, 727, 37
\bibitem[Band \&
Preece(2005)]{2005astro.ph..1559B} Band, D.~L., \& Preece, R.~D.\
2005, ArXiv Astrophysics e-prints, astro-ph/0501559
\bibitem[Ghirlanda et al.(2005)]{2005astro.ph..2186G} Ghirlanda, G.,
Ghisellini, G., \& Firmani, C.\ 2005, ArXiv Astrophysics e-prints,
astro-ph/0502186 - GGF05.
\bibitem[Lamb et al., 2004]{Lambetal}Lamb, D. Q., Donaghy, T. Q. \& Graziani, C. A  New Astronomy
Review 48, 459–464 (2004).
\bibitem[Nakar \&
Piran(2004)]{2004astro.ph.12232N} Nakar, E., \& Piran, T.\ 2004,
ArXiv Astrophysics e-prints, astro-ph/0412232


\end{thebibliography}
\end{document}